\documentclass[preprintnumbers,amsmath,amssymb,superscriptaddress]{revtex4}
\usepackage{epsfig}
\usepackage{psfrag}
\usepackage{amsfonts}
\usepackage{graphicx}
\usepackage{dcolumn}
\usepackage{bm}

\begin{document}


\title{The noncommutativity of the static and homogeneous limit of the axial chemical potential in chiral magnetic effect }

\author{Bo  Feng}
\affiliation{
School of Physics, Huazhong University of Science and Technology, Wuhan 430074, China
}

\author{De-fu Hou}
\affiliation{Institute of Particle Physics and Key Laboratory of Quark and Lepton Physics (MOE), Huazhong Normal University, Wuhan 430079, China}

\author{Hai-cang Ren}
\affiliation{Institute of Particle Physics and Key Laboratory of Quark and Lepton Physics (MOE), Huazhong Normal University, Wuhan 430079, China}
\affiliation{Physics Department, The Rockefeller University, 1230 York Avenue, New York, New York 10021-6399, USA}

\author{Shuai  Yuan}
\affiliation{
	School of Physics, Huazhong University of Science and Technology, Wuhan 430074, China
}


\date{\today}

\begin{abstract}

We study the noncommutativity of different orders of zero energy-momentum limit pertaining to the axial chemical potential in the chiral magnetic effect. While this noncommutativity issue originates from the pinching singularity at one-loop order, it cannot be removed by introducing a damping term to the 
fermion propagators. The physical reason is that modifying the propagator alone would violate the axial-vector Ward identity and as a result a modification of the longitudinal component of the axial-vector vertex is required, which contributes to CME. The pinching singularity with free fermion propagators was then taken over by the 
singularity stemming from the dressed axial-vector vertex. We show this mechanism by a concrete example.  Moreover, we proved in general the vanishing CME in the limit order that the static limit was taken prior to the homogeneous limit in the light of Coleman-Hill theorem for a static external magnetic field. For the opposite limit that the homogeneous limit is taken first, we show that the nonvanishing CME was a consequence of the nonrenormalizability of chiral anomaly for an arbitrary external magnetic field. 

\end{abstract}

\maketitle


\section{Introduction}

The collective macroscopic behavior of chiral matter subject to an external magnetic field or a vorticity field, by the interplay with chiral anomaly,  could manifest in anomalous transport phenomena. For instance, a vector current along the magnetic field could be induced in response to the magnetic field in the presence of a chirality imbalance, which is known as chiral magnetic effect (CME) \cite{CME1, CME2, CME3}.  It is of great interests to the phenomenology in the relativistic heavy ion collisions \cite{CMEinSTAR, CMEinALICE, CMEinCMS}, as well as in the condensed matter systems, such as the Weyl and Dirac semimetals \cite{Son_Spivak,CMEinSemimetals_1,CMEinSemimetals_2}.  It is believed that the charge separation observed in the correlation of final hadrons in noncentral heavy ion collisions and the negative magentoresistance observed in some semimetals are the consequences of the CME. Because of the noisy background, however, the CME in heavy ion collisions remains controversial and the intensive investigations are ongoing \cite{Kharzeev_Liao_Voloshin_Wang, CME_taskforce, Wang_Zhao, Liu_Huang}. 

With the chirality imbalance proxied by a constant axial chemical potential $\mu_5$, the chiral magnetic current in a constant magnetic field $\bf {B}$ takes the simple form \cite{CME2,CME3}
\begin{equation}
{\bf J} = \eta\frac{e^2}{2\pi^2}\mu_5{\bf B},
\label{classicalcme}
\end{equation}
with $\eta$ a factor associated with color and flavor degrees of freedom. In the reality of heavy ion collisions, however, both magnetic field and chirality imbalance are inhomogeneous and time dependent, and (\ref{classicalcme}) serves an 
approximation for slowly varying $\mu_5$ and $\bf {B}$. With an arbitrary magnetic field and arbitrary axial chemical potential, the chiral magnetic current in momentum 
representation reads
\begin{equation}
J^i(q+k) = \eta\frac{e^2}{2\pi^2}{\cal G}^{ij0}(q,k)\mu_5(k)A^j(q),
\label{kernel}
\end{equation}
where ${\cal G}^{\mu\nu\rho}(q,k)$ is proportional to the AVV three-point functions with one of the photon vertices and the axial-vector vertex 
bearing incoming 4-momenta $q$ and $k$. The current in the form (\ref{classicalcme}) corresponds to its infrared limit, $q\to 0$ and $k\to 0$, but this limit is subtle at a nonzero temperature.

At a constant $\mu_5$, i.e., $k=0$, ${\cal G}^{ij0}(q,0)$ can be parametrized as 
\begin{equation}
{\cal G}^{ij0}(q,0)=-iF(q)\epsilon^{ijk}q^k. \label{CMEcurrent_1}
\end{equation}
It was first found in \cite{CME3} that the limits ${\bf q}\to 0$ and $q^0\to 0$ do not commutate, 
i.e., $\lim_{{\bf q}\rightarrow 0}\lim_{q^0\rightarrow 0}F(q)=1$ and $\lim_{q^0\rightarrow 0}\lim_{{\bf q}\rightarrow 0}F(q)=1/3$. This noncomutativity in different order of limits was later confirmed by the calculations with Pauli-Villars \cite{HOU_LIU_REN} and lattice regularizations \cite{BO_HOU_REN}. Note that, however, the calculations with a proper regularization give rise to different results, i.e.,  $\lim_{{\bf q}\rightarrow 0}\lim_{q^0\rightarrow 0}F(q)=0$ and $\lim_{q^0\rightarrow 0}\lim_{{\bf q}\rightarrow 0}F(q)=-2/3$.  In the Pauli-Villars regularization scheme, for example, the extra contribution making the difference comes entirely from the regulator term.

The authors of \cite{Satow_Yee} provided a resolution to this noncomutativity problem by considering the interacting chiral fermion system. Specifically, it amounts to replace the free fermion propagator with a dressed fermion propagator incorporating a damping term in its self-energy. As a result, because of the finite relaxation time, the pinching singularity underlying the noncomutativity in different order of limits disappears. The authors therefore found that in both order of limits the form factor $F(0)=1$. This result is also consistent with the calculation in the strong coupling regime using the AdS/CFT correspondence \cite{ADS/CFT_Yee}, where the limit $q\to 0$ is unambiguous. An interesting discussion of this noncommutativity problem in the framework of chiral kinetic theory including Berry curvature\cite{Berrycurvature1, Berrycurvature2, Berrycurvature3, Berrycurvature4} was presented in \cite{Anatomy_CME}, where a new contribution called magnetization current to CME was identified  and attributed to the removal of the discontinuity of the CME conductivity in different order of limits. 

Coming to the infrared limit $k=(k^0, {\bf k})\to 0$, the explicit one-loop calculation under the Pauli-Villars regularization in \cite{HOU_LIU_REN} reveals the following noncomutativity. If the static limit of the chiral imbalance is prior to its homogeneity limit, 
\begin{equation}
\lim_{{\bf k}\to 0}\lim_{{k^0}\to 0}{\cal G}^{ij0}(q,k)=0,
\label{order1}
\end{equation}
for a static magnetic field in the homogeneous limit, i.e. $q=(0,{\bf q})\to 0$. In the opposite order, if the homogeneity limit is prior to the static limit   
\begin{equation}
\lim_{k^0\to 0}\lim_{{\bf k}\to 0}{\cal G}^{ij0}(q,k)=i\epsilon^{ijk}q^k, \label{order2}
\end{equation}
for arbitrary $q$. The latter order of limit gives rise to the chiral magnetic current 
\begin{equation}
{\bf J} = -\eta\frac{e^2}{2\pi^2}\mu_5{\bf B}, \label{nonclassicalcme}
\end{equation}
which differs from (\ref{classicalcme}) by a sign. The sign difference, however, cannot be detected with parity-even signal such as charge separation in heavy ion collisions or magnetoresistance in Weyl/Dirac metals. The authors of \cite{HOU_LIU_REN} related (\ref{order1}) to the vector Ward identity and (\ref{order2}) to the anomalous axial-vector Ward identity. Both identities go beyond one-loop order suggesting that the noncommutativity of the infrared limit persists to all orders.

As will be shown in the next section, the noncommutativity associated to the axial-vector vertex at one-loop level stems from the same pinching singularity as that underlying 
the noncommutativity of the photon vertex. A natural question that arises is why the dressed propagator fails to smear the difference between (\ref{order1}) and (\ref{order2}), and its answer together with related analysis occupy the rest of this work. Briefly speaking, a Ward identity links the longitudinal component of a vertex function with 
respect to the 4-momentum transfer to the self-energy function of the fermion propagator attached to it. Therefore it is inconsistent to modify a fermion propagator alone.  In case of the vertex of the magnetic field, only the transverse component contributes so the inconsistency does not manifest. This, however, is not the case with the axial-vector vertex. While the limit order (\ref{order1}) projects out the transverse component of the vertex and the inconsistency does not contribute, the opposite order of limits (\ref{order2}) does pick up the longitudinal component and the modification of the vertex function cannot be ignored. Through a subset of diagrams contributing to CME with the recipe \cite{Satow_Yee} of the modified propagator, we shall demonstrate that the  role of the pinching singularity of free propagators is taken over by the new infrared singularity of the modified axial-vector vertex and the difference between the two orders of limits (\ref{order1}) and (\ref{order2}) remains.

The rest of the paper is organized as follows: in Section II we shall present a one-loop calculation in order to elucidate the role of pinching singularity in the noncommutativity issue at the axial-vector vertex. A recapitulation of the Ward identity arguments in \cite{Golkar_Son} in the light of the Coleman-Hill's theorem and the nonrenormalization theorem of anomaly will be presented in Section III for self-containedness.  In Section IV, a concrete example is given for demonstrating the mechanism of the failure of the dressed fermion propagator in the noncommutativity issue. Section V concludes the paper.  Except for section III.A, we shall work in the framework of the 
closed-time path (CTP) Green's functions which is detailed in Appendix A. Throughout the paper, we will work with Minkowski metric and four vectors represented by $x^\mu=(x^0,{\bf x}), {q^\mu=(q^0, {\bf q})}$ with $q^0$ the energy.

\section{One-loop analysis}

\begin{figure}
	\includegraphics[height=4cm]{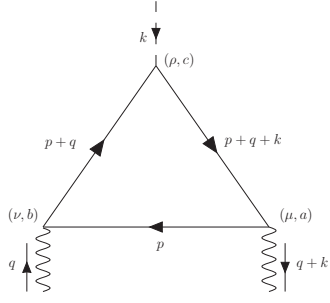}
	\caption{ The AVV triangle diagram. There is a second diagram with the photon four-momenta and polarization as well as CTP indices interchanged. } \label{Fig-1}
\end{figure}

In this section, we shall calculate the contribution to CME from the one-loop AVV three-point function in terms of CTP Green functions at a nonzero temperature in order to exhibit the role of the pinching singularity. The CTP Green's functions\cite{CTP} are generated by a path integral whose action is the integration of the classical Lagrangian along a closed time path that consists of a forward branch and a backward branch. The quantum field operator in CTP is denoted by $\phi_a(x)$ with $a=1,2$ labels the forward and backward branches. For more details about CTP Green's fuctions, see appendix A and Ref. \cite{CTP}. The amplitude of Fig.1 in CTP formalism reads
\begin{align}
 \nonumber {\cal G}^{\mu\nu\rho}_{abc}(q,k)=-ie^2\int\frac{d^4p}{(2\pi)^4}{\rm Tr}&\Big[\Gamma^\mu\eta_a S(p+q+k)\Gamma^{\rho 5}\eta_cS(p+q)\Gamma^\nu\eta_b S(p)\\
 &+\Gamma^\mu\eta_a S(p+q+k)\Gamma^\nu\eta_b S(p+k)\Gamma^{\rho 5}\eta_cS(p)\Big],
\end{align}
where the trace Tr(...) was extended to Dirac and CTP indices.  The subscripts in ${\cal G}^{\mu\nu\rho}_{abc}$ are CTP indices each taking values 1 and 2, which are projected by the $2\times 2$ matrices 
\begin{equation}
\eta_1=\left(\begin{array}{cc} 1 & 0\\
0 & 0\\ \end{array}\right), 
\ \ \ \ \ \hbox{and} 
\qquad \eta_2=\left(\begin{array}{cc} 0 & 0\\
0 & 1\\ \end{array}\right).
\label{projection}
\end{equation}
The bare vertices and propagators take the form
\begin{equation}
\Gamma^\mu=\left(\begin{array}{cc}
\gamma^\mu & 0\\
0 & -\gamma^\mu\\
\end{array}\right), \ \ \ \ \ \ \ \ \ \ \Gamma^{\mu 5}=\left(\begin{array}{cc}
\gamma^\mu\gamma^5 & 0\\
0 & -\gamma^\mu\gamma^5\\
\end{array}\right),\label{barevertex}
\end{equation}
with the negative sign taking into account of the reversed time integration along the backward branch, and
\begin{equation}
S(p)=\left(\begin{array}{cc}
S_{11} & S_{12}\\
S_{21} & S_{22}\\
\end{array}\right),	\label{ctpS}
\end{equation}
with
\begin{align}
	S_{11}(p) =& \frac{i}{p\!\!\!/+i0^+}-2\pi p\!\!\!/ \Big[ \theta(-p_0)+\epsilon(p_0)f_F(p_0)\Big]\delta(p^2),\nonumber\\
	S_{12}(p) =& -2\pi p\!\!\!/ \epsilon(p_0)f_F(p_0)\delta(p^2),\nonumber\\
	S_{21}(p) =& -2 \pi p\!\!\!/ \epsilon(p_0)\Big[f_F(p_0)-1\Big]\delta(p^2),\nonumber\\
	S_{22}(p) =& \frac{-i}{p\!\!\!/-i0^+}-2\pi p\!\!\!/\Big[ \theta(-p_0)+\epsilon(p_0)f_F(p_0)\Big]\delta(p^2),
\end{align}
for massless fermions with $f_F(x)=1/(e^{\beta x}+1)$ the Fermi distribution function at temperature $T=1/\beta$. We also defined $\theta(x)=1$ if $x>0$ and vanishing otherwise as well as $\epsilon(x)=\pm1$ for positive and negative $x$ respectively.  The AVV three-point function underlying retarded responses of the CME current to the magnetic field and axial chemical potential was obtained by restricting the electric current operator within the forward branch and summing up the rest CTP indices, i.e., 
\begin{align}
\nonumber {\cal G}^{ij0}_R(q,k)=&\sum_{b,c}{\cal G}^{ij0}_{1bc}(q,k)\\
\nonumber=&-ie^2\int\frac{d^4p}{(2\pi)^4}{\rm Tr}\gamma^i{\Big[} 
S_R(p+q+k)\gamma^0\gamma^5S_R(p+q)\gamma^jS_C(p)+ S_C(p+q+k)\gamma^0\gamma^5S_A(p+q)\gamma^jS_A(p)\\
\nonumber &+S_R(p+q+k)\gamma^0\gamma^5S_C(p+q)\gamma^jS_A(p)
+S_C(p+q+k)\gamma^jS_A(p+k)\gamma^0\gamma^5S_A(p)\\
&+S_R(p+q+k)\gamma^jS_C(p+k)\gamma^0\gamma^5S_A(p)
+S_R(p+q+k)\gamma^j{S}_R(p+k)\gamma^0\gamma^5S_C(p){\Big]},\label{retardedresponse}
\end{align}	
where we have switched to the physical representation of CTP formalism and the trace tr(...) extends to Dirac indices only. The physical representation can be obtained from (\ref{ctpS}) by an orthogonal transformation given in the appendix A and we have the retarded and advanced fermion propagators, and the correlator given by
\begin{align}
&S_R(p)=\frac{i}{p\!\!\!/+i0^+\gamma^0},\\
&S_A(p)=\frac{i}{p\!\!\!/-i0^-\gamma^0},\\
&S_C(p)=2\pi p\!\!\!/(1-2f_F(p))\delta(p^2).	
\end{align}
Substituting the KMS relation $S_C(p)=[1-2f_F(p^0)][S_R(p)-S_A(p)]$,  we have
\begin{align}
\nonumber {\cal G}^{ij0}_R(q,k)=&-ie^2\int\frac{d^4p}{(2\pi)^4}{\rm Tr}\gamma^i\\
\nonumber &\times\Big\{	f_F(p^0)\Big[S_R(p+q+k)\gamma^0\gamma^5S_R(p+q)\gamma^jS_R(p)+S_R(p+q+k)\gamma^jS_R(p+k)\gamma^0\gamma^5S_R(p)\\
\nonumber &-S_R(p+q+k)\gamma^0\gamma^5S_R(p+q)\gamma^jS_A(p)-S_R(p+q+k)\gamma^j{\cal S}_R(p+k)\gamma^0\gamma^5S_A(p)\Big]\\
\nonumber&-f_F(p^0+q^0+k^0)\Big[S_A(p+q+k)\gamma^0\gamma^5S_A(p+q)\gamma^jS_A(p)+S_A(p+q+k)\gamma^jS_A(p+k)\gamma^0\gamma^5S_A(p)\\
\nonumber&-S_R(p+q+k)\gamma^0\gamma^5S_A(p+q)\gamma^jS_A(p)-S_R(p+q+k)\gamma^jS_A(p+k)\gamma^0\gamma^5S_A(p)\Big]\\
\nonumber&+f_F(p^0+q^0)\Big[S_R(p+q+k)\gamma^0\gamma^5S_R(p+q)\gamma^jS_A(p)-S_R(p+q+k)\gamma^0\gamma^5S_A(p+q)\gamma^jS_A(p)\Big]\\
&-f_F(p^0+k^0)\Big[S_R(p+q+k)\gamma^j S_A(p+k)\gamma^0\gamma^5S_A(p)- S_R(p+q+k)\gamma^jS_R(p+k)\gamma^0\gamma^5S_A(p)\Big]\Big\}.
\end{align}	
For a static magnetic field, i.e. $q=(0,{\bf q})$,  the above equation is simplified to
\begin{align}
\nonumber {\cal G}^{ij0}_R(q,k)=&-ie^2\int\frac{d^4p}{(2\pi)^4}{\rm Tr}\gamma^i\\
\nonumber &\times{\Big\{} f_F(p^0)\Big[S_R(p+q+k)\gamma^0\gamma^5S_R(p+q)\gamma^jS_R(p)+S_R(p+q+k)\gamma^jS_R(p+k)\gamma^0\gamma^5S_R(p)\Big]\\
\nonumber &+[f_F(p^0+k^0)-f_F(p_0)]\Big[S_R(p+q+k)\gamma^0\gamma^5S_A(p+q)\gamma^jS_A(p)+	S_R(p+q+k)\gamma^jS_R(p+k)\gamma^0\gamma^5S_A(p)\Big]\\
&-f_F(p^0+k^0)\Big[S_A(p+q+k)\gamma^0\gamma^5 S_A(p+k)\gamma^jS_A(p)+S_A(p+q+k)\gamma^jS_A(p+k)\gamma^0\gamma^5S_A(p)\Big]\Big\}.
\label{1loopRes}
\end{align}

The noncommunitativity in the orders of limits $k\to 0$ stems from the pinching singularity of the terms of the structure $S_R\gamma^0\gamma^5S_A$. If $k^0\to 0$ at
a nonzero ${\bf k}$, the poles of $S_R$ and $S_A$ on the complex $p^0$ plane are separate and prefactor $[f_F(p^0+k^0)-f_F(p^0)]$ vanishes. The vanishing prefactor 
in this case renders the pinching  not contributing when the limit ${\bf k}\to 0$ was taken afterwards and we end up with
\begin{equation}
\lim_{{\bf k}\to 0}\lim_{k^0\to 0}{\cal G}^{ij0}_R(q,k)=e^2\int\frac{d^4p}{(2\pi)^4}f_F(p^0)\frac{\partial}{\partial p^0}{\rm Tr}\gamma^i\Big[S_R(p+q)\gamma^5\gamma^jS_R(p)-S_A(p+q)\gamma^5\gamma^jS_A(p)\Big],
\end{equation}
where the identities
\begin{equation}
\frac{\partial}{\partial p^0}S_{R(A)}(p)=iS_{R(A)}(p)\gamma^0S_{R(A)}(p),
\end{equation}
are employed.  

In the opposite order of limit, starting with ${\bf k}=0$, we use the relation \footnote{This relation can be deduced from the Ward identities 
(\ref{RAaxialwardidentities}) with free propagators and vertex or can be verified explicitly.} 
\begin{equation}
S_R(p+k)\gamma^0\gamma^5S_A(p)=S_R(p+k)\gamma^0\gamma^5S_R(p)+\frac{i}{k^0}[S_R(p)-S_A(p)]\gamma^0.
\end{equation}
Then in the limit $k^0\to 0$, the poles of $S_R(p+k)$ and $S_R(p)$ of the first term coalesce below the path of $p^0$ integration. As one is free to deform the 
integration away from the doubles poles, this term does not contribute when multiplied by the vanishing prefactor $[f_F(p^0+k^0)-f_F(p^0)]$. On the other hand, the 
$1/k^0$ of the second term together with the prefactor generates a nonzero contribution $-i(\partial f_F(p^0)/\partial p^0)[S_R(p)-S_A(p)]$ in the limit $k^0\to 0$ and we obtain that
\begin{equation}
\lim_{k^0\to 0}\lim_{{\bf k}\to 0}{\cal G}^{ij0}_R(q,k)=e^2\int\frac{d^4p}{(2\pi)^4}\frac{\partial}{\partial p^0}{\Big\{}f_F(p^0){\rm Tr}\gamma^i 
\Big[S_R(p+q)\gamma^5\gamma^jS_R(p)-S_A(p+q)\gamma^5\gamma^jS_A(p)\Big]{\Big\}}.\label{onelooporder2}
\end{equation}

It appears that the pinching singularity can be removed by introducing damping term in the fermion propagator \cite{Satow_Yee, Damping_1, Damping_2}, i.e.
\begin{align}
&S_R(p)\to{\cal S}_R(p)=\frac{i}{p\!\!\!/+\frac{i}{\tau}\gamma^0},\\
&S_A(p)\to{\cal S}_A(p)=\frac{i}{p\!\!\!/-\frac{i}{\tau}\gamma^0}.
\label{dressed}	
\end{align}
Then the poles of ${\cal S}_R(p+k)$ and ${\cal S}_A(p)$ will never coalesce in any orders of limit $k\to 0$ and the noncommutativity disappears. While this approach of 
smearing the noncommutativity of limits works for the magnetic vertex, it violates the axial Ward identity in our case as will be demonstrated in subsequence sections.

\section{A general analysis}

In this section, we shall prove that the noncommutativity issue at the axial-vector vertex persists to all orders of perturbations.  In one order of limits where the static limit is taken prior to the spatial homogeneity limit, we shall follow the argument of Coleman-Hill theorem\cite{Coleman_Hill}. In the opposite order of limits where the spatial homogeneity limit is taken first, we shall extend the diagramatical technique employed to derive the anomalous Ward identity at zero temperature to CTP Green's functions. In what follows, the kernel ${\cal G}^{ij0}(q,k)$ in (\ref{kernel}) under the two orders of limits, $\lim_{{\bf k}\to 0}\lim_{k^0\to 0}$ and $\lim_{k^0\to 0}\lim_{{\bf k}\to 0}$, will be analyzed non-perturbatively. It is convenient to revert to conventional notation pertaining the three-point function underlying the axial anomaly, i.e. $\Lambda^{\mu\nu\rho}(q_1, q_2)$ with $q_1$ and $q_2$ the 4-momenta of two outgoing photons. In terms of Fig. 1, we then have  ${\cal G}^{\mu\nu\rho}(q,k)=\Lambda^{\mu\nu\rho}(q+k,-q)$ with $k=q_1+q_2$ and $q=-q_2$. 

\subsection{The order of limit $\lim_{{\bf k}\to 0}\lim_{k^0\to 0}$}

Assuming a static magnetic field, i.e. $q^0=0$,  the first limit $k^0\to 0$ renders all external momenta static and the system is in thermal equilibrium. The general three-point function ${\cal G}^{ij0}(q,k)$ could be evaluated by 
means of the Matsubara formalism. Let 
\begin{equation}
\Gamma^{ij}({\bf q}_1,{\bf q}_2)\equiv\lim_{k^0\to 0}{\Lambda}^{ij0}(q+k,-q).
\end{equation}
We have, in the limit ${\bf k}\to 0$,

{\it Theorem:} $\Gamma^{ij}({\bf q,\bf q})={\cal O}(|{\bf q}|^2)$ as ${\bf q}\rightarrow 0$.

{\it Proof:}  
Consider a particular diagram of $\Gamma^{ij}({\bf q}_1,{\bf q}_2)$ where the photon vertex of outgoing momentum $(0,{\bf q}_2)$ is located in a fermion loop of $n$ photon 
vertices and the loop momentum ${\bf p}$. Setting ${\bf q}_2=0$, the diagram with the outgoing photon vertex removed from this loop (with $n-1$ photon vertices left over) is called a progenitor of $\Gamma^{ij}$ following Coleman and Hill \cite{Coleman_Hill}, and is denoted by $G^i({\bf q}_1)$. We have 
\begin{equation}
G^i({\bf q}_1)=\int\frac{d^3{\bf p}}{(2\pi)^3}F^i({\bf p},{\bf q}_1). \label{progenitoramplitude}
\end{equation}
where the summation over the Matsubara energy running through the loop has been included in $F^i({\bf p},{\bf q}_1)$. In the presence of the gauge invariant regulator, e.g., Pauli-Villars regulator, shifting integration momentum is legitimate, we have
\begin{equation}
G^i({\bf q}_1)=\int\frac{d^3{\bf p}}{(2\pi)^3}F^i({\bf p}+{\bm\delta},{\bf q}_1), \label{momentumshift}
\end{equation}
with $\bm\delta$ an arbitrary constant vector. Consequently, all terms of the Taylor expansion of (\ref{momentumshift}) in nonzero powers of $\bm\delta$ vanish. To the 
linear power
\begin{equation}
\delta_j\int\frac{d^3{\bf p}}{(2\pi)^3}\frac{\partial}{\partial p_j}F^i({\bf p},{\bf q}_1)=0.
\end{equation}
Since $\bm\delta$ is nonzero and its direction is arbitrary, one has
\begin{equation}
\int\frac{d^3{\bf p}}{(2\pi)^3}\frac{\partial}{\partial p_j}F^i({\bf p},{\bf q}_1)=\Gamma^{ij}({\bf q}_1,0)=0.
\end{equation}
The last equality follows from the observation that taking derivative with respect to the loop momentum amounts to restore the removed photon vertex of zero outgoing 
momentum because of
\begin{equation}
\frac{\partial}{\partial p_j}S(p)=S(p)\gamma^j S(p).
\end{equation}
Repeating the same argument for the progenitor with the photon vertex of outgoing momentum $(0, {\bf q}_1)$, we end up with
\begin{equation} 
\Gamma^{ij}(0, {\bf q}_2)=0.
\end{equation}
Consequently $\Gamma^{ij}(0,0)=0$ and
\begin{align}
\left.  \frac{\partial \Gamma_{ij}}{\partial q_k}\right|_{{\bf q}=0}
	=\left.\frac{\partial}{\partial q_{1k}}\Gamma_{ij}({\bf q}_1,0)\right|_{{\bf q}_1=0}+\left.\frac{\partial}{\partial q_{2k}}\Gamma_{ij}(0,{\bf q}_2)\right|_{{\bf q}_2=0}=0. \label{derivative}
\end{align}
The theorem is thereby proved. The validity of this theorem requires at least two independent external spatial momenta which is the case for a three point function. Also, the presence of axial-vector vertices is not essential for its proof, unlike the anomaly nonrenormalization theorem presented in subsection. III.B. As a corollary of the theorem, the chiral magnetic current under a constant magnetic field vanishes with this order of limits. 


As an illustration of the afore proved theorem, we shall calculate explicitly the contribution of the one-loop triangle diagram to the chiral magnetic current shown in Fig. 1 with 4-momenta $q=(0,{\bf q})$ and $k=(0,{\bf k})$
\begin{align}
\nonumber\Lambda^{ij 0}(q+k,-q)=&ie^2\mu_5\frac{1}{\beta}\sum_n\int\frac{d^3 {\bf p}}{(2\pi)^3}{\Big\{}{\rm tr}\left[\gamma^iS(p+q+k|0)\gamma^0\gamma^5S(p+q|0)\gamma^j S(p|0) \right]\\
&-\sum_s C_s{\rm tr}\left[\gamma^i S(p+q+k|M_s)\gamma^0\gamma^5 S(p+q|M_s)\gamma^j S(p|M_s)\right]+((q+k)\leftrightarrow -q, i\leftrightarrow j){\Big\}},\label{onelooptriangle}
\end{align}
with 
\begin{equation}
S(p|m)=\frac{i}{p\!\!\!/-m},
\end{equation}
the free propagator for quarks. Note that, we have regularized the amplitude by Pauli-Villars regularization with
$\sum_sC_s=1$ and $M_s\rightarrow\infty$ after the integration. In the limit ${\bf k}\rightarrow 0$,  we have
\begin{align}
	\nonumber \Lambda^{ij 0}(q,-q)=&ie^2\mu_5\frac{1}{\beta}\sum_n\int\frac{d^3 {\bf p}}{(2\pi)^3}{\Big\{}{\rm tr}\left[\gamma^i S(p+q|0)\gamma^0\gamma^5S(p+q|0)\gamma^j S(p|0) \right]\\
	&-\sum_s C_s{\rm tr}\left[\gamma^i S(p+q|M_s)\gamma^0\gamma^5 S(p+q|M_s)\gamma^j S(p|M_s)\right]+(q\leftrightarrow -q,i\leftrightarrow j){\Big\}}.
\end{align}
Using the identity
\begin{align}
 S(p|m)(i\gamma^0\gamma^5)S(p|m)
=-\frac{2m}{p^2-m^2}S(p|m)\gamma^0\gamma^5-\frac{\partial S(p|m)}{\partial p^0}\gamma^5,
\end{align}
we find that 
\begin{align}
	\nonumber\Lambda^{ij 0}(q,-q)=&-e^2\mu_5\frac{1}{\beta}\sum_n\int\frac{d^3 {\bf p}}{(2\pi)^3}\left\{{\rm tr}\left[\gamma^i \frac{\partial S(p+q|0)}{\partial p^0}\gamma^5 \gamma^j S(p|0) \right]\right.\\
	&\left.-\sum_s C_s{\rm tr}\left[\gamma^i \left(\frac{2M_s}{(p+q)^2-M_s^2}S(p+q|M_s)\gamma^0\gamma^5+\frac{\partial S(p+q|M_s)}{\partial p^0}\gamma^5\right)\gamma^j S(p|M_S)\right]+(q\leftrightarrow -q,i\leftrightarrow j)\right\},
\end{align}
where the derivative with respect to $p^0$ is evaluated at the Matsubara frequency $p^0=i(2n+1)\pi T$.
Since the integral is properly regularized, one can shift the momentum in the last terms in parenthesis by $p\rightarrow p+q$ and has
\begin{align}
\nonumber\Lambda^{ij 0}(q,-q)=&-e^2\mu_5\frac{1}{\beta}\sum_n\int\frac{d^3{\bf p}}{(2\pi)^3}\left\{\frac{\partial }{\partial p^0}\Xi^{ij}(p,p+q|0)-\sum_sC_s\frac{\partial }{\partial p^0}\Xi^{ij}(p,p+q|M_s)\right.\\
&-\sum_sC_s\left.\left[\frac{2M_s}{p^2-M_s^2}\Theta^{ij}(p,p+q|M_s)+\frac{2M_s}{(p+q)^2-M_s^2}\Theta^{ji}(p+q,p|M_s)\right]\right\},\label{linearmu5}
\end{align}	
with 
\begin{equation}
\Xi^{ij}(p,p+q|m)={\rm tr}[\gamma^i S(p|m)\gamma^5\gamma^j S(p+q|m)],
\end{equation}
and
\begin{equation}
\Theta^{ij}(p,p+q|m)={\rm tr}[\gamma^i S(p|m)\gamma^0\gamma^5\gamma^j S(p+q|m)].
\end{equation}

Although the first term on the RHS of (\ref{linearmu5}) is a total derivative, its contribute does not vanish at nonzero temperature due to the distribution function that emerges from converting the Matsubara summation to a contour integral. We have
\begin{align}
\nonumber \frac{1}{\beta}\sum_n\frac{\partial }{\partial p^0}\Xi^{ij}(p,p+q|0)\bigg\rvert_{p^0=i(2n+1)\pi T}=&  -\frac{1}{2\pi i}\oint_C dp^0\frac{\partial }{\partial p^0}\Xi^{ij}(p,p+q|0)f_F(p^0)\\
\nonumber =&\frac{1}{2\pi i}\oint_C dp^0\Xi^{ij}(p,p+q|0)\frac{\partial f_F(p^0)}{\partial p^0}\\
=&2 q_k\epsilon^{ijk} \frac{\beta}{\pi }\oint_C dp^0\frac{p^0}{[(p^0)^2-{\bf p}^2]^2}\frac{ e^{\beta p^0}}{\left(1+e^{\beta p^0}\right)^2}+{\cal O}({\bf q}^2). \label{surfaceterm}
\end{align}
In the last step,  we have taken the limit ${\bf q}\rightarrow 0$ and kept only the linear term in  $\bf q$. Carrying out the integrals,  we obtain

\begin{equation}
 -e^2\mu^5\frac{1}{\beta}\sum_n\int\frac{d^3 {\bf p}}{(2\pi)^3}\frac{\partial }{\partial p^0}\Xi^{ij}(p,p+q|0)\bigg\rvert_{p^0=i(2n+1)\pi T}=-i\frac{e^2}{2\pi^2} \mu^5q_k\epsilon^{ijk}. \label{unregularizedcontribution}
\end{equation}

The contribution of the Pauli-Villars regulators on the RHS of (\ref{linearmu5}) can be calculated without employing the Matsubara formalism even at nonzero temperature because of the large regulator mass $M_s$ in the quark propagators. Then the total derivative vanishes and we have 
\begin{align}
\nonumber &e^2\mu^5\sum_sC_s\int\frac{d^4 p}{(2\pi)^4}\left[\frac{2M_s}{p^2-M_s^2}\Theta^{ij}(p,p+q|M_s)+\frac{2M_s}{(p+q)^2-M_s^2}\Theta^{ji}(p+q,p|M_s)\right]\\
\nonumber =&-16ie^2\mu_5q_k \epsilon^{ijk}\sum_sC_sM_s^2\int\frac{d^4 p}{(2\pi)^4}\frac{1}{(p^2-M_s^2)^2[(p+q)^2-M_s^2]}+{\cal O}({\bf q}^2)\\
=&  i\frac{e^2}{2\pi^2}\mu_5q_k \epsilon^{ijk}, \label{regulatorcontribution}
\end{align}
where we used the fact that $\sum_sC_s=1$. The contribution of the Pauli-Villars regulators (\ref{regulatorcontribution}) cancels that of the unregularized part (\ref{unregularizedcontribution}) in the one-loop triangle diagram. Therefore, for a properly regularized AVV three-point function,  the CME current vanishes in the order of limits $\lim_{{\bf k}\to 0}\lim_{k^0\to 0}$ for a static external magnetic field up to the linear order in its spatial momentum $\bf q$. This is exactly the expected result from the theorem above.

\subsection{The order of limit $\lim_{k^0\to 0}\lim_{{\bf k}\to 0}$}

\begin{figure}
	\includegraphics[height=5cm]{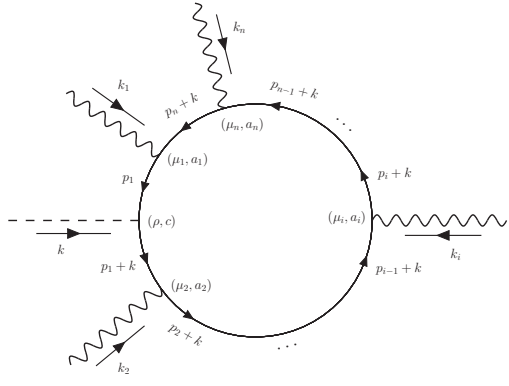}
	\caption{The fermion loop with one axial-vector vertex and $n>1$ vector vertices. } \label{Fig-1}
\end{figure}

We shall consider in this subsection the other order of limit, i.e., taking the spatial homogeneity limit prior to the static limit in the axial-vector vertex. Let's start with a fermion loop, denoted by $\Gamma$, with one axial-vector vertex and $n>1$ vector vertices as shown in Fig.2. In this loop,  the axial-vector field with incoming momentum $k$ was inserted between vertices $\mu_1$  and $\mu_2$. By convention, the additional momentum $k$ running around the fermion loop from the axial-vector vertex would exit at vertex $\mu_1$. We have in CTP formalism
\begin{align}
\nonumber &\Gamma^{\mu_1\rho\mu_2\cdots\mu_n}_{a_1 c a_2\dots a_n}(k, k_1, \cdots, k_n)\\
=&-(-ie)^n\int\frac{dp_1}{(2\pi)^4}{\rm Tr}\Big[\Gamma^{\rho 5}\eta_cS(p_1)\Gamma^{\mu_1}\eta_{a_1}S(p_n+k)\Gamma^{\mu_n}\eta_{a_n}S(p_{n-1}+k) \cdots S(p_3+k)\Gamma^{\mu_3}\eta_{a_3}S(p_2+k)\Gamma^{\mu_2}\eta_{a_2}S(p_1+k)\Big],
\end{align}
where $\Gamma^{\mu 5}$ and $S(p)$, given in (\ref{barevertex}) and (\ref{ctpS}), are the bare axial-vector vertex and quark propagator, respectively. 
Taking the divergence with respect to the axial-vector vertex and employing the identity
\begin{equation}
S(p+k)(-ik_\mu)\Gamma^{\mu 5}\eta_a S(p)=\eta_a\gamma^5 S(p)+S(p+k)\gamma^5\eta_a,
\label{tree}
\end{equation}
for massless fermions, we have
	\begin{align}
	\nonumber	&ik_\rho\Gamma^{\mu_1\rho\mu_2\cdots\mu_n}_{a_1 c a_2\dots a_n}(k, k_1, \cdots, k_n)\\
		=&(-ie)^n\int\frac{dp_1}{(2\pi)^4}{\rm Tr}\Big[\Gamma^{\mu_2}\eta_{a_2}\eta_c\gamma^5 \left(S(p_1)-S(p_1+k)\right)\Gamma^{\mu_1}\eta_{a_1}S(p_n+k)\Gamma^{\mu_n}\eta_{a_n}S(p_{n-1}+k) \cdots S(p_3+k)\Gamma^{\mu_3}\eta_{a_3}S(p_2+k)
		\Big].\label{divergencemu1mu2}
	\end{align}
Likewisely, if the axial-vector vertex was inserted between vertices $\mu_2$ and $\mu_3$,  we will have
	\begin{align}
\nonumber &	ik_\rho\Gamma^{\mu_1\mu_2\rho\mu_3\cdots\mu_n}_{a_1  a_2 c a_3\dots a_n}(k, k_1, \cdots, k_n)\\
	=&(-ie)^n\int\frac{dp_1}{(2\pi)^4}{\rm Tr}\Big[\Gamma^{\mu_2}\eta_{a_2}\eta_c\gamma^5 S(p_1)\Gamma^{\mu_1}\eta_{a_1}S(p_n+k)\Gamma^{\mu_n}\eta_{a_n} S(p_{n-1}+k)\cdots S(p_3+k)\Gamma^{\mu_3}\eta_{a_3}\left(S(p_2)-S(p_2+k)\right)
	\Big].\label{divergencemu2mu3}
\end{align}
Therefore the first term in (\ref{divergencemu1mu2}) will be canceled by the second term in (\ref{divergencemu2mu3}). Similar cancellations take place between terms from other pairs of diagrams with adjacent insertions of axial-vector vertices. Summing over all $n$ insertions, we end up with
	\begin{align}
	\nonumber &\sum_{i=1}^{n}	ik_\rho\Gamma^{\mu_1\mu_2\cdots\mu_{i}\rho\mu_{i+1}\cdots\mu_n}_{a_1  a_2 \cdots a_i c a_{i+1}\cdots a_n}(k, k_1, \cdots, k_n)\\
	\nonumber =&-(-ie)^n\int\frac{dp_1}{(2\pi)^4}{\rm Tr}\Big[\Gamma^{\mu_2}\eta_{a_2}\eta_c\gamma^5 S(p_1+k)\Gamma^{\mu_1}\eta_{a_1}S(p_n+k)\Gamma^{\mu_n}\eta_{a_n} S(p_{n-1}+k)\cdots S(p_3+k)\Gamma^{\mu_3}\eta_{a_3}S(p_2+k)\\
	&-\Gamma^{\mu_2}\eta_{a_2}\eta_c\gamma^5 S(p_1)\Gamma^{\mu_1}\eta_{a_1}S(p_n)\Gamma^{\mu_n}\eta_{a_n} S(p_{n-1})\cdots S(p_3)\Gamma^{\mu_3}\eta_{a_3}S(p_2)
	\Big],
\end{align}
where the number $i$ in the summation  is modulated in $n$. Shifting the integration variable from $p_1$ to $p_1+k$ in the second term, if it is legitimate, the two remaining terms cancel. As shown by Adler and Bardeen \cite{Adler_Barteen}, the only circumstance that invalidates this momentum shift was the diagram with $n=2$. While the legitimacy of the momentum shifting can be restored by a UV regulator, say Pauli-Villars regulator, the regulator mass invalidates (\ref{tree}) and gave rise to the anomalous Ward identity
\begin{equation}
-i(k_1+k_2)_\rho\Gamma^{\rho\mu_1\mu_2}_{ca_1a_2}(k_1,k_2)=\left\{\begin{array}{c}
\frac{e^2}{4\pi^2}\epsilon^{\mu_1\mu_2\alpha\beta}k_{1\alpha}k_{2\beta}, \ \ \ \ \ c=a_1=a_2=1,\\
-\frac{e^2}{4\pi^2}\epsilon^{\mu_1\mu_2\alpha\beta}k_{1\alpha}k_{2\beta}, \ \ \  c=a_1=a_2=2,\\
0, \ \ \ \ \ \ \ \ \  \ \ \ \ \ \ \ \ \ \ \ \ \ \ \ \ \ \ {\rm otherwise}.\\
\end{array}\right.
\end{equation}
Consequently, the chiral magnetic current with this order of limits takes the form of (\ref{nonclassicalcme}) \footnote{Incidentally,  a radiative correction to the anomalous Ward identity stemming from a three-loop diagram with the two photons coming from the triangle diagram rescattered was found in \cite{Ansel'm}. The same diagram contributes to the CME current at zero temperature but the contribution vanishes in the limit considered here at a nonzero temperature for a static magnetic field. See \cite{Rediative_correction} for details.}. This results is in contrast to the null result in the order of limit, $\lim_{{\bf k}\to 0}\lim_{k^0\to 0}$,  which is only valid in linear order of a small magnetic field momentum.



\section{A concrete example}

In this section, we shall show a concrete example which demonstrates that dressing the fermion propagators could not smear the noncommutativity of the order of the spatial homogeneity limit and the static limit in the axial-vector vertex. 
One important lesson we learn is that in the AVV three-point function, in order to preserve the (axial-) vector Ward identity at the (axial-) vector 
vertex not only the fermion propagators but also the (axial-) vector vertex should be dressed accordingly. 

In CTP formulation, the Ward identities satisfied by the vector and axial-vector vertices are
\begin{equation}
{\cal S}(p^\prime)\sigma_1(-iq_\mu)\Gamma^{\mu}(p^\prime,p)\sigma_1{\cal S}(p)={\cal S}(p)-{\cal S}(p^\prime),\label{vectorwardidentity}
\end{equation}  
and
\begin{equation}
{\cal S}(p^\prime)\sigma_1(-iq_\mu)\Gamma^{\mu 5}(p^\prime,p)\sigma_1{\cal S}(p)=\gamma_5{\cal S}(p)+{\cal S}(p^\prime)\gamma_5,\label{axialwardidentity}
\end{equation}  
in the physical representation, 
where ${\cal S}(p)$ is the full fermion propagator carrying both spinor and CTP indices, $\Gamma^\mu,\Gamma^{\mu5}$ are the amputated full vector and 
axial-vector vertex functions with $q=p^\prime-p$ the momentum flowing into it, and $\sigma_1$ is the Pauli matrix with respect to CTP indices, which always accompanies the CTP Green functions in physical representations. As shown, the Ward identity ties the longitudinal component of the vertex function to the propagator. Any dressing of the latter has to be reflected in the longitudinal component of the former. In case of the triangle diagrams underlying CME, while the longitudinal component of the photon vertex contributes to neither order of limit of the response to magnetic field, the longitudinal components of the axial-vector vertex does contribute to the order of limit in subsection III.B. The mechanism of removing the ambiguity of 
the orders of limit with respect to the magnetic field does not contradict to the vector Ward identity but fails here respect to the axial-vector Ward identity. 

In terms of the explicit form of the fermion propagator and axial-vector vertex function
\begin{equation}
		{\cal S}(p)=\left(\begin{array}{cc} 0 & {\cal S}_A(p)\\
			                {\cal S}_R(p) & {\cal S}_C(p)\\ \end{array}\right), 
\label{physical}
\end{equation}
and
\begin{equation}
		\Gamma^{\mu5}(p',p)=\left(\begin{array}{cc} 0 & \Gamma_A^{\mu5}(p',p)\\
			                \Gamma_R^{\mu5}(p',p) & \Gamma_C^{\mu5}(p',p)\\ \end{array}\right), 
\label{vertex}
\end{equation}
the Ward identity (\ref{axialwardidentity}) becomes
\begin{subequations}
\begin{equation}
{\cal S}_A(p^\prime)(-iq_\mu)\Gamma_A^{\mu 5}(p^\prime,p){\cal S}_A(p) = \gamma^5{\cal S}_A(p)+{\cal S}_A(p^\prime)\gamma^5,
\end{equation} 
\begin{equation}
{\cal S}_R(p^\prime)(-iq_\mu)\Gamma_{R}^{\mu 5}(p^\prime,p){\cal S}_{R}(p) = \gamma^5{\cal S}_R(p)+{\cal S}_R(p^\prime)\gamma^5,
\end{equation}
\begin{equation}
{\cal S}_R(p^\prime)(-iq_\mu)\Gamma_{RA}^{\mu 5}(p^\prime,p){\cal S}_A(p) = \gamma^5{\cal S}_A(p)+{\cal S}_R(p^\prime)\gamma^5,
\end{equation}\label{RAaxialwardidentities}
\end{subequations}
where
\begin{equation}
\Gamma_{RA}^{\mu 5}(p',p)\equiv \frac{\Gamma_C^{\mu 5}(p',p)+[1-2f_F(p^{0\prime})]\Gamma_A^{\mu 5}(p',p)-[1-2f_F(p^0)]\Gamma_R^{\mu 5}(p',p)}
{2[f_F(p^0)-f_F(p^{0\prime})]},
\label{GammaRA}
\end{equation}
and the KMS relation
\begin{equation}
{\cal S}_C(p)=[1-2f_F(p^0)][{\cal S}_R(p)-{\cal S}_A(p)],\label{KMSinCTP}
\end{equation}
is employed.
To the zeroth order, ${\cal S}_A(p)$ and ${\cal S}_R(p)$ are given by free propagators and $\Gamma_A^{\mu 5}(p^\prime,p)=\Gamma_R^{\mu 5}(p^\prime,p)
=\gamma^\mu\gamma^5$ and $\Gamma_C^{\mu 5}=0$. Consequently, $\Gamma_{RA}^{\mu 5}=\gamma^\mu\gamma^5$. Note that, in general, a full axial-vector vertex function $\Gamma^{\mu 5}(p',p)$ contains eight components with respect to the CTP indices. The external line of the axial-vector vertex, however, was restrained to occur only on the forward branch. Consequently,  the number of components of the full axial-vector vertex reduces to four and it can be written by a $2\times 2$ form as in (\ref{vertex}). For the  specific definitions of the components in (\ref{vertex}), see the appendix A.

Let's consider the AVV three-point function shown in Fig. 3 with full propagators and a modified axial-vector vertex, and bare vector vertices. The retarded CME kernel implied by this diagram has the form
\begin{align}
\nonumber {\cal G}^{ij0}_R(q,k)=-ie^2\int\frac{d^4p}{(2\pi)^4}{\rm Tr}\gamma^i \Big[
\nonumber&  {\cal S}_R(p+q+k)\Gamma_R^{05}(p+q+k,p+q){\cal S}_R(p+q)\gamma^j{\cal S}_C(p)\\
\nonumber+&  {\cal S}_R(p+q+k)\Gamma_C^{05}(p+q+k,p+q){\cal S}_A(p+q)\gamma^j{\cal S}_A(p)\\
\nonumber+&  {\cal S}_C(p+q+k)\Gamma_A^{05}(p+q+k,p+q){\cal S}_A(p+q)\gamma^j{\cal S}_A(p)\\
\nonumber+&  {\cal S}_R(p+q+k)\Gamma_R^{05}(p+q+k,p+q){\cal S}_C(p+q)\gamma^j{\cal S}_A(p)\\
\nonumber+&  {\cal S}_C(p+q+k)\gamma^j{\cal S}_A(p+k)\Gamma_A^{05}(p+k,p){\cal S}_A(p)\\
\nonumber+&  {\cal S}_R(p+q+k)\gamma^j{\cal S}_R(p+k)\Gamma_C^{05}(p+k,p){\cal S}_A(p)\\
\nonumber+&  {\cal S}_R(p+q+k)\gamma^j{\cal S}_C(p+k)\Gamma_A^{05}(p+k,p){\cal S}_A(p)\\
+&  {\cal S}_R(p+q+k)\gamma^j{\cal S}_R(p+k)\Gamma_R^{05}(p+k,p){\cal S}_C(p)\Big].\label{fullretardedresponse}
\end{align} 
Employing the KMS relation (\ref{KMSinCTP}), we can rewrite the retarded CME kernel for a static magnetic field in terms of retarded and advanced propagators
\begin{align}
\nonumber {\cal G}^{ij0}_R(q,k)=-ie^2\int\frac{d^4p}{(2\pi)^4}{\rm Tr}\gamma^i\Big\{ 
\nonumber&  f_F(p^0)\Big[{\cal S}_R(p+q+k)\Gamma_R^{05}(p+q+k,p+q){\cal S}_R(p+q)\gamma^j{\cal S}_R(p)\\
\nonumber+&             {\cal S}_R(p+q+k)\gamma^j{\cal S}_R(p+k)\Gamma_R^{05}(p+k,p){\cal S}_R(p)\Big]\\
\nonumber+& \Big[f_F(p^0+q^0)-f_F(p^0)\Big]\Big[{\cal S}_R(p+q+k)\Gamma_{RA}^{05}(p+q+k,p+q){\cal S}_A(p+q)\gamma^j{\cal S}_A(p)\\
\nonumber-&             {\cal S}_R(p+q+k)\gamma^j{\cal S}_R(p+k)\Gamma_{RA}^{05}(p+k,p){\cal S}_A(p)\Big]\\
\nonumber -&  f_F(p^0+k^0)\Big[{\cal S}_A(p+q+k)\Gamma_A^{05}(p+q+k,p+q){\cal S}_A(p+q)\gamma^j{\cal S}_A(p)\\
+&{\cal S}_A(p+q+k)\gamma^j{\cal S}_A(p+q)\Gamma_A^{05}(p+q,p){\cal S}_A(p)\Big]\Big\},\label{step1}
\end{align}
with $q=(0,{\bf q})$ and $k=(k^0,{\bf k})$, in parallel to the one-loop expression (\ref{1loopRes}).  
Considering a homogeneous axial chemical potential, i.e. $k=(k^0,{\bf 0})$, and taking the divergence at the axial-vector vertex, we have 
\begin{align}
\nonumber k^0{\cal G}^{ij0}_R=e^2\int\frac{d^4p}{(2\pi)^4}&\Big\{
          f_F(p^0){\rm Tr}\gamma^i\gamma^5[{\cal S}_R(p+q)\gamma^j{\cal S}_R(p)-S_A(p+q)\gamma^j{\cal S}_A(p)]\\
-&  f_F(p^0+k^0){\rm Tr}\gamma^i\gamma^5[{\cal S}_R(p+q+k)\gamma^j{\cal S}_R(p+k)-{\cal S}_A(p+q+k)\gamma^j{\cal S}_A(p+k)]
\Big\}.	\label{Totalderivative}
\end{align}	
where the Ward identities  (\ref{RAaxialwardidentities}) were employed. It follows that
\begin{equation}
\lim_{k^0\to 0}\lim_{{\bf k}\to 0}{\cal G}^{ij0}_R(q,k)=e^2\int\frac{d^4p}{(2\pi)^4}\frac{\partial}{\partial p^0}\Big\{f_F(p^0){\rm Tr}\gamma^i\Big[{\cal S}_R(p+q)\gamma^5\gamma^j{\cal S}_R(p)-{\cal S}_A(p+q)\gamma^5\gamma^j{\cal S}_A(p)\Big]\Big\}, \label{step2}
\end{equation}
in parallel to (\ref{onelooporder2}). In a proper regularization, for instance the Pauli-Villars regularization, the momentum shift will be legitimate and the two terms cancel out. The regulator term will contribute to the nonzero current (\ref{order2}).

\begin{figure}
	\includegraphics[height=6cm]{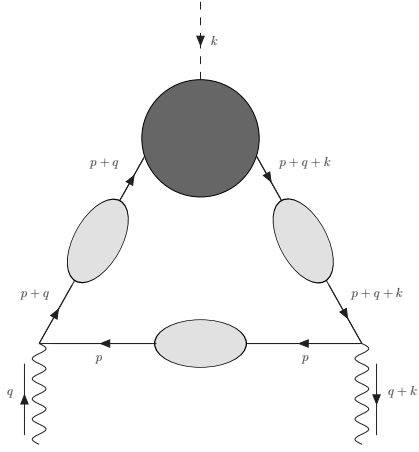}
	\caption{Subdiagrams with a modified axial-vector vertex and full fermion propagators, but bare vector vertices. There is a second set of subdiagrams with the photon four-momenta and polarization indices interchanged. } \label{Fig-1}
\end{figure}

To be specific, let us consider the dressed propagator (\ref{dressed}). It follows from (\ref{RAaxialwardidentities}) and (\ref{GammaRA}) for $p^\prime-p=(k^0,{\bf 0})$ that
\begin{align}
\nonumber\Gamma_R^{05}(p^\prime,p) &=\Gamma_A^{05}(p^\prime,p)=\gamma^0\gamma^5,\\
\nonumber\Gamma_{RA}^{05}(p^\prime,p) &=\gamma^0\gamma^5+\frac{2i}{k^0\tau}\gamma^0\gamma^5,\\
\Gamma_C^{05}(p^\prime,p) &= \frac{4i}{k^0\tau}[f_F(p^0)-f_F(p^0+k^0)]\gamma^0\gamma^5.
\end{align}  
Consequently, the role of the pinching singularity with the bare propagator and axial-vector vertex is taken over by the $1/k^0$ singularity of the dressed axial-vector 
vertex function $\Gamma_{RA}^{05}(p^\prime,p)$ in the limit of $k^0\to 0$. In another word, it is precisely this singularity that facilitates the reduction from (\ref{step1}) to (\ref{step2}).  

We have to emphasize that Fig.3 only represents a subset of diagrams underlying the radiative corrections to the AVV triangle whose integrand adds up to a total derivative with 
respect to the energy running through the Fermion loop and thereby contributing to the non-renormalization of the anomaly independent of other diagrams. The dressed axial-vector vertex in Fig.3 excludes the diagrams in Fig.4, which adds another term to the RHS of (\ref{axialwardidentity}) because of the anomaly \cite{Adler, Adler_Barteen}. The photon vertices 
in Fig.3 remains undressed. Dressing the photon vertices amounts to introduce more bare photon vertices with internal photon lines attached to them and the logics from (\ref{physical}) to (\ref{Totalderivative}) is expected to carried through. On the other hand, even with dressed photon vertices, the diagrams included in Fig.3 do not cover all radiative corrections to 
AVV triangle. An example not included in Fig.3 can be found in Ref. \cite{Ansel'm, Rediative_correction}. Nevertheless, the selection of the subset of diagrams in Fig.3 demonstrates that modifying  the fermion propagator with damping does not remove ambiguity of the infrared limit of the 4-momentum pertaining the axial chemical potential.

\begin{figure}
	\includegraphics[height=5cm]{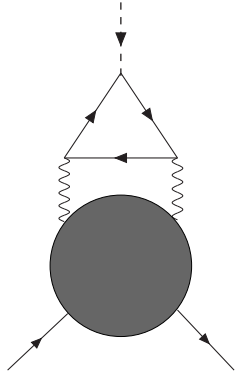}
	\caption{An axial-vector vertex not considered in Fig. 3.} 
\end{figure}

\section{Conclusions and Outlooks}

In this paper, we studied the noncommutativity of different orders of zero energy-momentum limit pertaining to the axial chemical potential in the chiral magnetic effect. The vanishing CME in the limit that the static limit was taken prior to the homogeneous limit was proved in general by an argument similar to the Coleman-Hill theorem for a static external magnetic field. For the opposite limit that the homogeneous limit is taken first the nonvanishing CME was a consequence of the nonrenormalizability of chiral anomaly. While the nonrenormalizability of chiral anomaly is valid for arbitrary external momenta, the Coleman-Hill theorem applies only in the limit ${\bf q}\rightarrow 0$ with ${\bf q}$ the spatial momentum of the external magnetic field. A possible caveats of the Coleman-Hill theorem is the infrared divergence at a nonzero temperature when more and more gluon loops 
are introduced. But the lattice simulation indicates a nonzero magnetic mass of gluons \cite{Lattice_Magnetic_Mass_1, Lattice_Magnetic_Mass_2, Lattice_Magnetic_Mass_3}, which serves an infrared cutoff. In addition, a recent calculation based on holography 
shows that such a noncommutativity stays in strongly coupled $N=4$ super Yang-Mills theory \cite{Yin_Lei}.

At one-loop level, the noncommutativity of different orders of zero momentum limit is originated from the pinching of the poles of the retarded and advanced fermion 
propagators at nonzero temperature. It is tentative to smear the pinching singularity by introducing a finite damping term to the fermion propagators and thereby remove the
noncommutativity between different orders of limits. The mechanism works for the photon vertex attached to the magnetic field and the infrared ambiguity is indeed removed. 
As we shown in this work, this approach does not work for the axial-vector vertex. The physical reason is that modifying the propagator alone would violate the vector and 
axial vector Ward identities, which requires a corresponding modification of the longitudinal component of the vector and axial vector vertex with respect to the 4-momentum 
transfer. For the vector vertex attached to the magnetic field, only the transverse component matters in either order of limits and the issue does not 
arise. For the axial vector vertex, the static limit after homogeneity limit picks up the longitudinal component and the modified vertex contributes. We demonstrate this point 
by a subset of diagrams contributing to CME and show explicitly that the difference between the two orders of limits remains as expected.

Our work at this stage is of theoretical value only. In view of the dynamical nature of the chirality imbalance in realistic heavy ion collisions, it is important to explore 
under which order of limits the constant $\mu_5$ approximation better describes the phenomenology of the chiral magnetic effect there.

\begin{acknowledgments}
The work of D-f H. and H-c R. is in part supported by the NSFC Grant  Nos. 11735007, 11890711.  B.F.'s work was supported in part by the NSF of China under Grant No. 11535005.

\end{acknowledgments}

\appendix

\section{Green functions and vertices in closed-time path formalism}

As the closed-time path (CTP) Green's function is less well-known than the Green's functions underlying the Feynman diagrams and Matsubara diagrams, we provide some background
behind the CTP formalism employed in this work. A systematic discussion of CTP formulation and its applications can be found in \cite{CTP}.

The CTP Green's functions are generated by a path integral whose action is the integration of the classical Lagrangian along a closed time path which consists of a forward branch, $\int_{-\infty}^\infty dt(...)$ and a backward branch, $\int_{\infty}^{-\infty} dt(...)$. The number of degrees of freedom is thereby doubled. The quantum field operator in CTP is denoted by $\phi_\alpha(x)$ with $\alpha=1,2$ labels
the forward and backward branches.
The time-ordering operator underlying the CTP Green's functions becomes the path-ordering along the closed time path, i.e. ordinary time ordering along the forward branch and anti-time ordering along the backward branch with the backward branch preceding the forward branch.   
The two-point Green function of  operators $A_\alpha(t_1)$ and $B_\beta(t_2)$ is defined as
\begin{equation}
D_{\alpha\beta}(t_1,t_2)=\langle T_p(A_\alpha(t_1)B_\beta(t_2)) \rangle,
\end{equation}
where $T_p$ enforces the path-ordering operator along the CTP contour and the dependence on the spatial and internal coordinates is suppressed for clarity.

It is convenient to write the two-point function in a $2\times 2$ matrix form, named as the single-time representation in \cite{CTP}, i.e.

\begin{equation}
D=\left(\begin{array}{cc}
D_{11},D_{12}\\
D_{21},D_{22}
\end{array}\right),\label{singletimerepre}
\end{equation}
with

\begin{subequations}
	\begin{equation}
	D_{11}(t_1,t_2)=\langle T(A(t_1)B(t_2))\rangle,
	\end{equation}
	\begin{equation}
	D_{12}(t_1,t_2)=\langle B(t_2)A(t_1)\rangle,
	\end{equation}
	\begin{equation}
	D_{21}(t_1,t_2)=\pm\langle A(t_1)B(t_2)\rangle,
	\end{equation}
	\begin{equation}
	D_{22}(t_1,t_2)=\langle{\tilde{T}} (A(t_1)B(t_2))\rangle,
	\end{equation}\label{singletimerepresentations}
\end{subequations}
where $T$ is the usual time-ordering operator, while $\tilde{T}$ is the anti-time-ordering operator. Whenever "$\pm$" or "$\mp$" shows up, the upper sign refers to bosons and the lower sign to fermions.  The four components in the matrix form (\ref{singletimerepre}) are not independent and satisfy the following identity
\begin{equation}
D_{11}+D_{22}=D_{12}+D_{21}.
\end{equation}
As shown in (\ref{singletimerepresentations}), once the operators are placed explicitly in the order with backward branch preceding the forward branch, 
the branch indices are removed since both $\phi_1(t)$ and $\phi_2(t)$ corresponds to the same Hilbert space operator. 
In particular, we have $T_p(\phi_1(t))=T_p(\phi_2(t))=\phi(t)$. 
The fermionic operator when multiplied by a variable can be treated as a bosonic operator and their CTP Green functions can be extracted after factorizing out the Grassmann 
variable pertaining to each fermionic operator.

The CTP Green functions are also defined with respect to "physical" field operators
\begin{align}
	\nonumber \phi_\Delta(t)&=\phi_1(t)-\phi_2(t),\\ \phi_c(t)&=\frac{1}{2}\left(\phi_1(t)+\phi_2(t)\right).\label{unitary}
\end{align}
Consequently, the physical representation (with $\Delta$ and $c$ indices) of the CTP Green function can be obtained from (\ref{singletimerepre}) by an orthogonal transformation, i.e.
\begin{equation}
D=V^{-1}{\cal D}V,\label{orthogonal}
\end{equation}
with
\begin{equation}
V=\frac{1}{\sqrt{2}}(1-i\sigma^2)=\frac{1}{\sqrt{2}}\left(\begin{array}{rr}
1 & -1\\
1 & 1\\ 
\end{array}\right), 
\ \ \ \ \ \ V^{-1}=\frac{1}{\sqrt{2}}\left(\begin{array}{rr}
1 & 1\\
-1 & 1\\ 
\end{array}\right),
\end{equation}
and we end up with
\begin{equation}
{\cal D}=\left(\begin{array}{cc}
0 & D_A\\
D_R & D_C\\ 
\end{array}\right),\label{physicalrepre}
\end{equation}
where
\begin{subequations}
	\begin{equation}
	D_{R}(t_1,t_2)=\theta(12)\langle [A(t_1),B(t_2)]_\mp\rangle\label{EqA},
	\end{equation}
	\begin{equation}
	D_{A}(t_1,t_2)=-\theta(21)\langle[A(t_1),B(t_2)]_\mp\rangle\label{EqB},
	\end{equation}
	\begin{equation}
	D_{C}(t_1,t_2)=\langle[A(t_1),B(t_2)]_\pm\rangle\label{EqC},
	\end{equation}
\end{subequations}
where $\theta(12)$ is the step function, which equals one if $t_1>t_2$ and vanishes otherwise, and $[\cdots]_\mp$ stands for commutator and anti-commutator. $D_R, D_A$ and $D_C$ are the retarded, advanced and correlation functions, respectively, and they satisfy the KMS relation
\begin{equation}
{D}_C(p)=[1-2f(p^0)][{D}_R(p)-{D}_A(p)],\label{KMS}
\end{equation}
with $f(p^0)=1/(e^{\beta p^0}\pm 1)$ the  Bose-Einstein or Fermi-Dirac distribution functions for the bosonic or fermionic fields, respectively. The same orthogonal transformation (\ref{orthogonal}) also converts the signature matrix of time integration along the CTP contour, $\sigma_3$ in the single time representation to $\sigma_1$ in the physical representation.

It is instructive to verify the structure of (\ref{physicalrepre}) directly from definition since the methodology can be readily extended to the three-point function considered in this work. The $\Delta\Delta$-component of (\ref{physicalrepre}) takes the form
\begin{align}
	\nonumber \langle T_p(A_\Delta(t_1)B_\Delta(t_2))\rangle=&\theta(12)\langle T_p(A_\Delta(t_1)B_\Delta(t_2))\rangle\pm\theta(21)\langle T_p(B_\Delta(t_2)A_\Delta(t_1)) \rangle\\
	=&\theta(12)\langle T_p(A_\Delta(t_1)\left(B_1(t_2)-B_2(t_2)\right))\rangle\pm\theta(21)\langle T_p(B_\Delta(t_2)\left(A_1(t_1)-A_2(t_1)\right))\rangle.
\end{align}
Looking at the first term on RHS, $t_2$ is the earliest moment, therefore, $B_1(t_2)$ should reside at the rightmost position and $B_2(t_2)$ at the leftmost position, i.e.
\begin{equation}
{\rm 1st\  term}=\theta(12)\langle (A_1(t_1)-A_2(t_1))B_1(t_2)\mp B_2(t_2)(A_1(t_1)-A_2(t_2))\rangle=\theta(12)\langle [A(t_1)-A(t_1), B(t_2)]_\mp\rangle=0.
\end{equation}
The same logic renders the second term vanish as well and we find $\langle T_p(A_\Delta(t_1)B_\Delta(t_2)) \rangle=0$. Next, let us consider the $\Delta c$-component. We have
\begin{equation}
\langle T_p(A_\Delta(t_1)B_c(t_2)) \rangle=\frac{1}{2}\theta(12)\langle T_p[(A_1(t_1)-A_2(t_1))(B_1(t_2)+B_2(t_2))]\rangle\pm\frac{1}{2}\theta(21)\langle T_p[(B_1(t_2)+B_2(t_2))(A_1(t_1)-A_2(t_1))]\rangle,
\end{equation}
where
\begin{equation}
{\rm 1st\ term}=\frac{1}{2}\theta(12)\langle (A_1(t_1)-A_2(t_1))B_1(t_2)\pm B_2(t_2)(A_1(t_1)-A_2(t_2))\rangle=\frac{1}{2}\theta(12)\langle[(A(t_1)-A(t_1)),B(t_2)]_\mp \rangle=0,
\end{equation}
and
\begin{equation}
{\rm 2nd\ term}=\frac{1}{2}\theta(21)\langle (B_1(t_2)+B_2(t_2))A_1(t_1)\mp A_2(t_1)(B_1(t_2)+B_2(t_2))\rangle=\theta(21)\langle [B(t_2),A(t_1)]_\mp\rangle,
\end{equation}
which gives rise to (\ref{EqB}). The same manipulation, when applied to the $c\Delta$-component, leads to (\ref{EqA}).

This reduction extends readily to $n$-point Green function with the recipe: 1) Write down all possible orders of the time variables by inserting the identity
\begin{equation}
1=\sum_{\cal P}\theta(p_1 p_2 ... p_{n-1} p_n)\label{permu},
\end{equation}
with $\theta(12...n)=\theta(12)\theta(23)...\theta((n-1)n)$ and the sum extends to all permutation of $1, 2, ..., n$.  2) Time order the operators inside $T_p(...)$ and 3) Remove the operators one by one from $T_p(...)$ according to their location in the forward or backward time branches. As an illustration, we consider the following three-point function
\begin{equation}
\langle T_p(A_\Delta(t_1)B_\alpha(t_2)C_\beta(t_3)) \rangle,
\end{equation}
with $\alpha, \beta=\Delta$ or $c$. The RHS of (\ref{permu}) consists six permutations of the time ordering. To have a nonzero $T_p$ product, the latest time  must be associated with a $c$-component. Consequently
\begin{equation}
\langle T_p(A_\Delta(t_1)B_\Delta(t_2)C_\Delta(t_3)) \rangle=0,
\label{DDD}
\end{equation}
For the $T_p$ product with one operator in $c$-component, we have
\begin{align}
	\nonumber \langle T_p(A_\Delta(t_1)B_\Delta(t_2)C_c(t_3)) \rangle=&\theta(312)\langle T_p(C_c(t_3)A_\Delta(t_1)B_\Delta(t_2)) \rangle\pm\theta(321)\langle T_p(C_c(t_3)B_\Delta(t_2)A_\Delta(t_1)) \rangle \\
	=&\theta(213)\langle [[C(t_3), A(t_1)]_\mp, B(t_2)]_- \rangle \pm\theta(321)\langle [[C(t_3), B(t_2)]_\mp, A(t_1)]_- \rangle,
\end{align}
and
\begin{align}
	\nonumber \langle T_p(A_\Delta(t_1)B_c(t_2)C_\Delta(t_3)) \rangle&=\pm\theta(213)\langle T_p(B_c(t_2)A_\Delta(t_1)C_\Delta(t_3)) \rangle+\theta(231)\langle T_p(B_c(t_2)C_\Delta(t_3)A_\Delta(t_1)) \rangle\\
	&=\pm\theta(213)\langle [[B(t_2), A(t_1)]_\mp, C(t_3)]_- \rangle+\theta(231)\langle [[B(t_2), C(t_3)]_\mp, A(t_1)]_- \rangle,
\end{align}
where, without loss of generality, we assumed that the operators $A(t), B(t)$ and $C(t)$ are simultaneously bosonic or femionic ones.

Now we are equipped to analyze the structure of the three point functions encountered in this work. While a three-point function has eight CTP components in general, not all components 
contribute to our case. If an operator underlying the three point-function couples to an external field, only $\Delta$-component of the operator contributes since the external field takes 
equal values on both time branches. For the AVV function, we associate $A(t_1)$ to the axial-vector current density coupling with the axial chemical potential, $B(t_2)$ to the 
electric current coupling with the magnetic field and $C(t_3)$ to the electric current to be measured. Thereby only two CTP components left over, i.e. $\Delta\Delta\Delta$- and 
$\Delta\Delta c$-components with the former one vanishing in according to (\ref{DDD}). Consequently
\begin{equation}
\langle T_p(A_\Delta(t_1)B_\Delta(t_2)C_c(t_3)) \rangle = \langle T_p(A_\Delta(t_1)B_\Delta(t_2)C_1(t_3)),
\end{equation}
which, upon applying the Wick theorem, gives rise to the retarded kernel (\ref{retardedresponse}) or (\ref{fullretardedresponse}).

Coming to the dressed axial vector vertex in (\ref{vertex}), we associate $A(t_1)$ to the axial vector current, and $B(t_2)$ and $C(t_3)$ fermionic fields. Only $A(t_1)$ couples to the external $\mu_5$, we are left with four CTP components, which can be packed in a $2\times 2$ matrix
\begin{equation}
\left(\begin{array}{cc} 
\langle T_p(A_\Delta(t_1)B_\Delta(t_2)C_\Delta(t_3)) \rangle & \langle T_p(A_\Delta(t_1)B_\Delta(t_2)C_c(t_3)) \rangle\\
\langle T_p(A_\Delta(t_1)B_c(t_2)C_\Delta(t_3)) \rangle & \langle T_p(A_\Delta(t_1)B_c(t_2)C_c(t_3)) \rangle\\ 
\end{array}\right),
\end{equation}
with the upper left element vanishes, resonating the structure of (\ref{vertex}).
Notice that the vertex function is obtained by amputating the two fermion legs and the amputation leaves the structure intact.

A comprehensive discussion of the general multi-point Green functions and vertices in the physical representation can be found in \cite{CTP}. 


\newpage 

\end{document}